\documentclass{article} 
\usepackage{general,times}
\usepackage{graphicx}
\usepackage[colorlinks = true,
            linkcolor = blue,
            urlcolor  = blue,
            citecolor = blue,
            anchorcolor = blue]{hyperref}
\usepackage{hyperref}
\usepackage{url}

\title{Deep High-Resolution Network for Low Dose X-ray CT Denoising}

\author{Ti Bai, Dan Nguyen, Biling Wang and Steve Jiang \\
Medical Artificial Intelligence and Automation (MAIA) Laboratory\\
Department of Radiation Oncology\\
University of Texas Southwestern Medical Centre  \\
Dallas, TX 75390, USA \\
Corresponding Author: Steve.Jiang@UTSouthwestern.edu
}
\begin{document}

\maketitle

\begin{abstract}
Low Dose Computed Tomography (LDCT) is clinically desirable due to the reduced radiation to patients. However, the quality of LDCT images is often sub-optimal because of the inevitable strong quantum noise. Inspired by their unprecedent success in computer vision, deep learning (DL)-based techniques have been used for LDCT denoising. Despite the promising noise removal ability of DL models, people have observed that the resolution of the DL-denoised images is compromised, decreasing their clinical value. Aiming at relieving this problem, in this work, we developed a more effective denoiser by introducing a high-resolution network (HRNet). Since HRNet consists of multiple branches of subnetworks to extract multiscale features which are later fused together, the quality of the generated features can be substantially enhanced, leading to improved denoising performance. Experimental results demonstrated that the introduced HRNet-based denoiser outperforms the benchmarked UNet-based denoiser in terms of superior image resolution preservation ability while comparable, if not better, noise suppression ability. Quantitative metrics in terms of root-mean-squared-errors (RMSE)/structure similarity index (SSIM) showed that the HRNet-based denoiser can improve the values from 113.80/0.550 (LDCT) to 55.24/0.745 (HRNet), in comparison to 59.87/0.712 for the UNet-based denoiser.
\end{abstract}

\section{Introduction}
X-ray computed tomography (CT) has been widely used in clinics to visualize the internal organs of patients for diagnostic purpose. However, radiation dose involved in the X-ray CT scans constitutes a potential health concern to human body, since it may induce genetic, cancerous, and other diseases. Consequently, following the As Low As Reasonably Achievable (ALARA) principle is of utmost importance in clinics when determining the radiation dose level of a CT scan.  One of the dominant ways to reduce the dose level is to decrease the exposure level of each projection angle. However, lower exposure level will inevitably induce stronger quantum noise into the CT image, possibly making accurate diagnosis, and organ contouring impossible, rendering decreased clinical value. Thus, denoising algorithm is essential for low dose CT (LDCT) image quality enhancement.

There has been a surge of research towards this direction, which roughly can be divided into three categories: projection domain-based denoising\cite{RN1,RN2}, image domain-based denoising\cite{RN3,RN4}, and regularized iterative reconstruction\cite{RN5,Rn6,RN7,RN8,RN9,RN10,RN11,RN12}. Because it is more convenient to access the CT images directly from the picture archive and communication system (PACS), image domain-based denoising algorithms are becoming more and more popular. In this work, we also devote our efforts on this category to enhance the image quality of LDCT.

Inspired by the unprecedent successes of convolutional neural network (CNN)-based deep learning (DL) techniques\cite{RN13} on various image processing\cite{RN14, RN15, RN16, RN17, RN18} and/or recognition\cite{RN19,RN20,RN21,RN22,RN23} tasks, they have also been widely introduced by the medicine community to denoise the CT images\cite{RN24,RN25,RN26,RN27,RN28,RN29,RN30,RN31}, achieving state-of-the-art denoising performance. For instance, Chen \textit{et al.} devised a deep CNN that is consisting of several repeating convolutional modules to learn a mapping from LDCT image to normal dose CT (NDCT) image\cite{RN25}. To facilitate a more efficient information flow and compensate the potential spatial resolution loss, they further proposed a new architecture by using the well-known residual learning scheme, called residual encoder-decoder (RED)\cite{RN25}, attaining better denoising performance. This performance improvement benefited from the more advanced architecture design. Indeed, it is well-recognized that neural network architecture is one of the dominant factors that affect the algorithm’s performance since it will directly affect the qualities of the features which are an abstract representation of the image. Moreover, the development history of CNN-based deep learning techniques in the past decade can be well-organized in terms of the evolvement of the associated architectures, such as from the original AlexNet\cite{RN19} that contains several convolutional layers and fully connected layers, to the VGG \cite{RN32} network by substantially increasing the network’s depth, to the Inception network \cite{RN33} by considering multiscale structures, to the well-known ResNet \cite{RN22} by proposing a residual connection to facilitate the information propagation, and to the feature pyramid network (FPN) \cite{RN34} by fusing the low-level and high-level features. There also exists many other architecture variants, such as SENet \cite{RN35}, DenseNet \cite{RN36}, and ResNeXt \cite{RN37}. As the network architectures evolving, the associated algorithm’s performance is also improved substantially, such as the top-5 error rate of the well-known ImageNet 1K dataset is reduced from 20.91\% (AlexNet) to 5.47\% (ResNeXt)\footnote{These data come from the model zoo provided by the official implementations of PyTorch. See https://pytorch.org/docs/stable/torchvision/models.html for more details.}.

Regarding our LDCT image denoising task, its goal is to suppress the noise effectively while preserve the resolution as much as possible since noise and resolution are two most important while combating metrics for the CT image quality evaluation. It is realized that high-level feature is the key to facilitate effective noise suppression by aggravating information from a large receptive field, while low-level feature is essential for nice resolution preservation. Therefore, high-quality low- and high-level feature extraction and effective fusion between them will be extremely important when design the CNN architectures for LDCT image denoising. Towards this technical line, U-Net \cite{RN38} might be one of the most well-known architectures by adding a skip connection to facilitate the fusion of the low-level features and the high-level features. Despite its great success in various tasks and datasets, when applying it on the LDCT image denoising task, people found that the U-Net architecture still leads to significant resolution loss, decreasing the clinical value of the denoised CT image. Therefore, there still exists room to further enhance the image quality of the denoised CT image by employing some more advanced network architecture that can extract and fuse both the low-level and high-level features more effectively.

In the natural image processing and analysis field, Sun \textit{et al.} \cite{RN39,RN40} proposed a high-resolution network (HRNet)-based representation learning method, attaining state-of-the-art performances among various computer vision tasks, such as human pose estimation, image classification, object detection, and segmentation. Specifically, HRNet uses multiple branches to extract multiscale features which are then internally fused together to facilitate the feature fusion. This nice architecture design can ensure a highly efficient low- and high-level features combination during the forward propagation process, and thus produce a high-quality feature for the downstream task.

To our best knowledge, this nice architecture has not been applied on denoising tasks. And we feel that HRNet is super suitable for medical image denoising tasks where high resolution is highly desirable to preserve the fidelity of anatomical structures. Realized this, in this work, we introduced HRNet into the medical image processing field, verified its superior denoising performance for the LDCT images. We hope this new introduced architecture can also inspire the other researchers in the medical image field to boost their tasks’ performance.

\section{Methods and Materials}
\subsection{Methods}
Let us first mathematically formulate our problem. Given a LDCT image $x\in\mathcal{R}^N$ whose noise component is represented as $\epsilon\in\mathcal{R}^N$, then the LDCT denoising task is to restore the underlying clean image $y\in\mathcal{R}^N$, where both the noisy and clean images are vectorized, $N$ denotes the number of the pixels. Without loss of generality, we can assume the noise is additive as:
\begin{equation}
\label{eq:noise model}
    x = y + \epsilon.
\end{equation}

In the DL’s viewpoint, problem~(\ref{eq:noise model}) can be regarded as an image to image translation problem that can be solved by learning a CNN $\Phi_W\left(x\right)$ who is parameterized by $W$. Given a pre-collected training dataset $\{\left(x_i,y_i\right)|i\in0,1,\cdots,M\}$ with a size of $M$, where $i$ indexes the training samples, the parameters W can be estimated by minimizing the following mean squared error (MSE)-based cost function:
\begin{equation}
\label{eq:cost function}
W=\arg\min_{W}\sum_{i}||\Phi_W(x_i) - y_i||_2^2.
\end{equation}

It should be noted that in practice, one can never access the real clean image that does not contain any noise since quantum noise is inevitable. Therefore, the clean image in the training dataset is usually replaced with a NDCT image $\tilde{y}$ that is also contaminated by mild noise. In this work, we also use this relaxed but practical experimental setting to train our network.

\begin{figure}
    \centering
    \includegraphics[width=\textwidth]{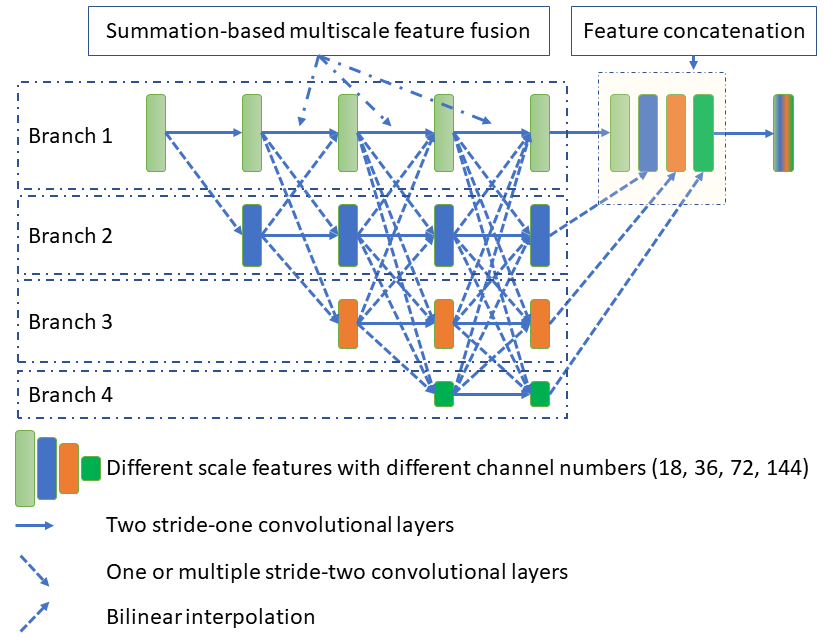}
    \caption{Illustration of the HRNet architecture. There are four different branches who are responsible to extract the features from different scales. During the forward propagation, the features from different scales would be fused together gradually. To enable the fusion process, as indicated by the dash line connections, the previous features will be first downsampled or upsampled depending on the relative size of the features such that the summation-based feature fusion is valid. Finally, all the features from different scales would be upsampled if necessary into the original scale and concatenated together, resulting in the final features. A predictor is attached as the last layer to output the denoised image. All the convolutional layers consist of three operators: convolution, instance normalization and nonlinear rectified unit. The final predictor consists of two operators: convolution and nonlinear rectified unit.}
    \label{fig:architecture}
\end{figure}

For completeness, we here also illustrate the architecture of the HRNet. 

As shown in Figure~(\ref{fig:architecture}), there are four different branches that are used to extract the features from different scales. The starting feature of each scale is calculated by applying a stride-two convolutional layer on the feature of the previous scale. All the features from different scales will be processed in a stage-by-stage fashion. Each stage contains two convolutional layers, as indicated by the solid arrow in Figure~(\ref{fig:architecture}). Before proceeding to the next stage, the features from different scales will be firstly fused together. In this work, this fusion process is conducted based on the feature summation. More specifically, if the size of the previous feature is larger than the current feature, such as fuse the features from branch 1 into the features from branch 3, one or multiple stride-two convolutional layers are used to downsample the previous feature such that it has same size as current feature. For example, if the previous feature has a size of $128\times128$ while the current feature has a size of $32\times32$, then two continuous stride-two convolutional layers are used. On the other hand, if the size of the previous feature is smaller than the current feature, bilinear interpolation is used to upsample the previous feature. Once we get all the features from different scales, they will be further concatenated together to generate the final feature. Note that those features having a smaller size will be firstly bilinearly interpolated into a size same as the original input image size. A predictor will be attached into the last layer to output the denoised image.

In this work, all the convolutional layers except the predictor consist of three operators: convolution operator, instance normalization \cite{RN41} operator and the rectified nonlinear unit (ReLU). The predictor is consisting of the convolution operator and the ReLU. The channel number for the features in same scale is the same, and detailed in Figure~(\ref{fig:architecture}). Since X-ray CT is gray-scaled image, the input and output channels are one.

\subsection{Materials}
\subsubsection{Training and validation datasets}
In this work, we used the public released AAPM Low Dose CT Grand Challenge dataset \cite{RN42} that consists of contrast-enhanced abdominal CT examinations for the model training (https://www.aapm.org/grandchallenge/lowdosect/). Specifically, NDCT projections were firstly acquired based on the routine clinical protocols in Mayo clinic by using the automated exposure control and automated tube potential selection strategies, such that the referenced tube potential is 120kV and the quality referenced effective exposure is 200mAs. Then Poisson noise was inserted into the above NDCT projection data to reach a noise level that corresponded to 25\% of the full dose, denoted as the LDCT projection data. Later, both the NDCT and the LDCT projection data were filtered back projected (FBP) into the image domain, producing the NDCT and LDCT images. The Challenge host provided twofour different reconstructions for both dose levels with two different slice thicknesses and two different kernels, i.e., $1\mathrm{mm}$ and $3\mathrm{mm}$., sharp kernel or not. In this work, we only used the $1\mathrm{mm}$ sharp reconstructions to test our model performance.

In the official training dataset, there are ten patient data. For qualitative evaluation, we further randomly split them into eight/two patient data, serving as the training/validation datasets, respectively. All the data involved in this work are in 2D slice. In total, there are 4800 and 1136 2D CT slice images in the training and validation datasets, respectively. 

For showcase purpose, we chose four representative slices in the validation datasets to demonstrate and compare the denoising performance. The first slice was used to check the denoising performance for the abdominal site which is the dominant site of the training dataset. The second slice corresponds to the liver which is one of the most important organs of human in the abdominal site. The third slice was selected to be in the lung region which was partially covered during the CT scan. The fourth slice contains a radiologists confirmed low-contrast lesion.

\subsubsection{Testing dataset}
We used a patient data from the testing dataset of the above Challenge to verify the model’s performance. Specifically, we firstly rebinned the raw helical-scanned projection data into fan-beam projection data with a slice thickness of $1\mathrm{mm}$ and a pixel size of $1\times1\mathrm{mm}^2$, which was then FBP reconstructed into the CT image domain. It is noted that the training dataset was reconstructed with the commercial implementation of the FBP algorithm, while our testing dataset was reconstructed with our homemade FBP algorithm. Thus, despite the same noise distribution in the projection domain, there exists certain domain gap in the image domain between this testing dataset and the training dataset due to different implementation details, such as different filter kernels.

\subsubsection{Training details}
The input and the target images were the LDCT and NDCT images, respectively. In details, both images would be firstly normalized by divided 2000 such that most of the image intensities fall into the range of 0-1. Note that the pixel values of the original CT images are CT values that were shifted by 1000 HU.

To enlarge the dataset, random translation (maximum 128 pixels along each axis), cropping, and rotation ($\pm 10^{\circ}$) were adopted to augment the training data samples. Zero-padding was employed when necessary to ensure the final image having a size of $512\times512$.

The Adam \cite{RN43} optimizer was used to minimize the cost function~(\ref{eq:cost function}), with hyperparameters $\beta_1=0.9$ and $\beta_2=0.999$. The algorithm was iterated by 100K iterations. The learning rate was initially set as to $1\times{10}^{-4}$, which was then reduced to be $1\times{10}^{-5}$ and $1\times{10}^{-6}$ at iterations 50K and 75K, respectively. The batch size was set as to 1. The PyTorch framework was used to train the network.

\subsubsection{Comparison studies}
For comparison, we also implemented a baseline network architecture: U-Net.

To be specific, in the encoder part, the initial output channel number in the input layer is 32. Continuous strid-two convolutional layers are applied to extract the image high-level features. The channel number is doubled per each downsampling resulted from the stride-two convolutional operators until reaching a feature number of 512. There are nine layers in the encoder part, leading to a bottleneck feature with a size of $1\times1$, suggesting that the associated receptive field covers the whole image that have a size of $512\times512$. In the decoder part, concatenation operation is adopted to fuse both the low-level and the high-level features. Th output channel number is one for the last output layer in the decoder part. Same as HRNet, the convolutional layer also contains three operators: convolution, instance normalization and ReLU.
All the other training details for the U-Net were the same as the above introduced HRNet.
 
\subsubsection{Evaluation metrics}
In this work, quantitative metrics in terms of the Root-Mean-Square-Error (RMSE) and the structure similarity (SSIM) \cite{RN44} are calculated to evaluate all the results. The RMSE is defined as:
$$\mathrm{RMSE}=\sqrt{||\bar{x}-\tilde{y}||_2^2},$$
where $\bar{x}$ and $\tilde{y}$ represented the evaluated image and the associated NDCT image if available.

The SSIM is defined as:
$$\mathrm{SSIM}(x,y)= \frac{(2\mu_{\bar{x}}\mu_{\tilde{y}} + a_1)(2\sigma_{\bar{x}}\sigma_{\tilde{y}} + a_2)}{(\mu_{\bar{x}}^2 + \mu_{\tilde{y}}^2 + a_1)(\sigma_{\bar{x}}^2 + \sigma_{\tilde{y}}^2 + a_2)},$$
where $\mu_{\bar{x}}$ and $\sigma_{\bar{x}}^2$ represents the mean value and the variance of the denoised image $\bar{x}$, same notations also apply for the NDCT image. We selected $a_1=1\times{10}^{-4}$ and $a_2=9\times{10}^{-4}$ to stabilize the calculation process.

Moreover, we also exploit the contrast-noise-ratio (CNR) to quantify the lesion detectability. The CNR is defined as follows:
$$\mathrm{CNR}=\frac{2|\mu_{\mathrm{fg}} - \mu_{\mathrm{bg}}|}{|\sigma_{\mathrm{fg}} - \sigma_{\mathrm{bg}}|}$$
where $\mathrm{\mu}_{\mathrm{fg}}$ and $\sigma_{\mathrm{fg}}$ correspond to the mean value and the standard deviation of the foreground (fg) region of interest (ROI), respectively. Same naming rules apply to the background (bg) ROI.

\subsubsection{Noise analysis} 
As mentioned above, the model is trained with paired LDCT-NDCT images, where both the input LDCT and the target NDCT images are contaminated by noise. More specifically, the target NDCT image contains realistic quantum noise (denoted as target noise hereafter), while the input LDCT image contains both the target noise inherited from the NDCT image as well as the extra added simulated noise (denoted as added noise hereafter). It will be interesting and valuable to analyze whether the trained denoiser can remove both the target noise and the added noise.

We first define the difference between the input LDCT and the output denoised image as removed noise. We use the cosine correlations between removed noise and added noise or target noise to characterize the composition of the removed noise. More specifically, since the target noise is coupled with the underlying clean image, and there may exist certain amount of structures in the removed noise, the cosine correlations are calculated based on the high frequency components, where the majority is the noise. In this work, the high frequency components are defined as those frequencies in the ranges of $\left[-\pi,-\frac{150}{256}\pi\right]$ and  $\left[\frac{150}{256}\pi,\pi\right]$. Moreover, we also calculate the projection lengths from the removed noise to the target noise and the added noise to quantify how much target noise and/or added noise are removed. 

As a showcase, all the slices from one patient in the validation dataset are used to analyze the noise correlations.

\section{Results}
\begin{figure}
    \centering
    \includegraphics[width=\textwidth]{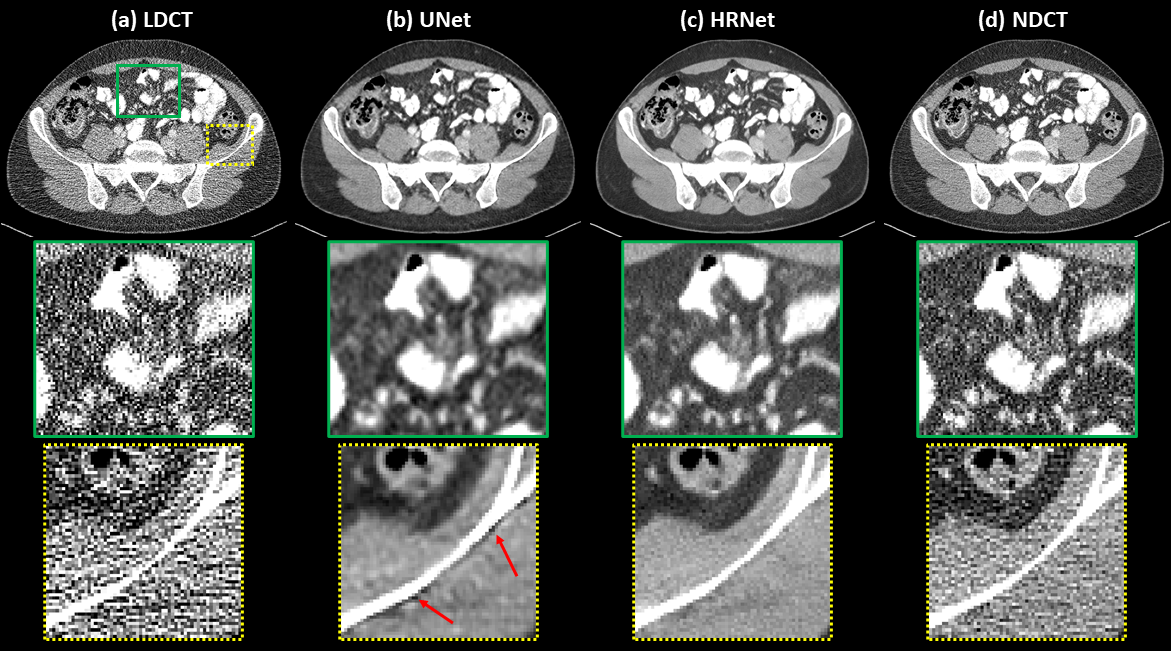}
    \caption{Denoised results for the slice selected from the abdominal site. (a) LDCT, (b) UNet, (c) HRNet, and (d) NDCT. Images in the second and the third rows are the zoomed-in views for the contents in the first row, corresponding to the region of interest covered by the solid green box and the dot yellow box, respectively. Display window: [-160, 240] HU.}
    \label{fig:abdomen}
\end{figure}

Figure~(\ref{fig:abdomen}) presents the denoised results for a slice in the validation dataset corresponding to the abdominal site. It is obvious that both denoisers can effectively suppress the noise, as shown in Figures~(\ref{fig:abdomen})(b) and~(\ref{fig:abdomen})(c). Compared to the NDCT image in Figure~(\ref{fig:abdomen})(d), we can find that the UNet-based denoiser produces an image with much inferior resolution, while the introduced HRNet-based denoiser can lead to an image (Figure~(\ref{fig:abdomen})(c)) with much better resolution than the UNet-based denoiser, while slightly inferior resolution than the NDCT image. These phenomena can be clearly observed from the zoomed-in views displayed in the second row of Figure~(\ref{fig:abdomen}). Besides, we can also find that the noise level of the image associated with the HRNet-based denoiser is much lower than that of the NDCT image. Another interesting finding is that the UNet-based denoiser leads to strong artifacts around the boundary of the bone, as indicated by the red arrows in the third row of Figure~(\ref{fig:abdomen}). By contrast, the HRNet-based denoiser can faithfully restore the underlying structures around that region.

\begin{figure}
    \centering
    \includegraphics[width=\textwidth]{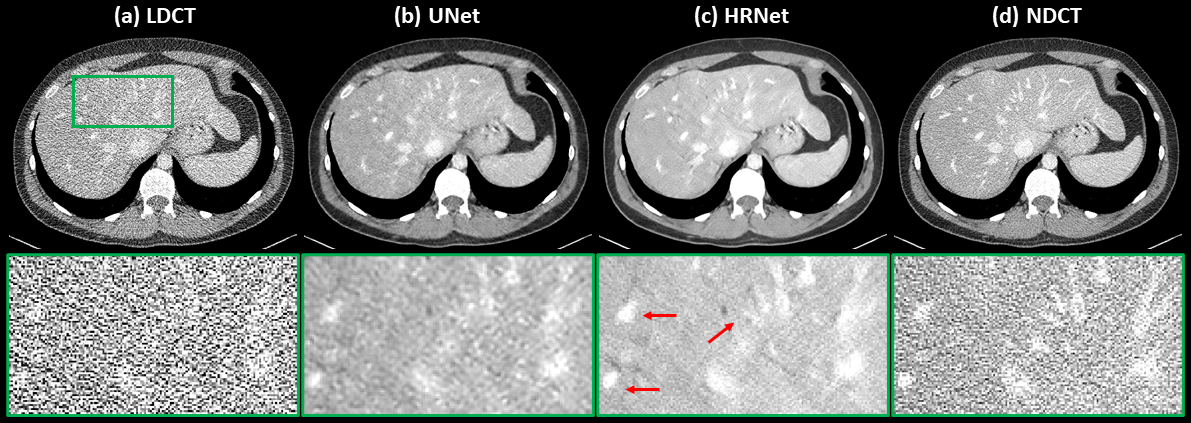}
    \caption{Denoised results for the slice corresponding to the liver organ. (a) LDCT, (b) UNet, (c) HRNet, and (d) NDCT. Images in the second row are the zoomed-in views for the contents in the first row, corresponding to the region of interest covered by the solid green box. Display window: [-160, 240] HU.}
    \label{fig:liver}
\end{figure}
Figure~(\ref{fig:liver}) demonstrates the denoised results of the liver organ. It can be seen that the strong noise in the LDCT image overwhelm the fine vessels in the liver. After applying any denoiser, the noise is suppressed effectively. Observation of the zoomed-in view in the second row of Figure~(\ref{fig:liver}) reveals that the HRNet-based denoiser produces an image that possesses more details, thereby suggesting higher resolution, than the UNet-based denoiser. Moreover, the noise in the image associated with the HRNet-based denoiser is also weaker than that of the UNet-based denoiser, indicating stronger denoising ability of the HRNet. Compared to the NDCT, as indicated by the red arrows, we can find that some of the vessels in the image with respect to the HRNet can be even more easily distinguished than those in the NDCT image which also suffers from the quantum noise. 

\begin{figure}
    \centering
    \includegraphics[width=\textwidth]{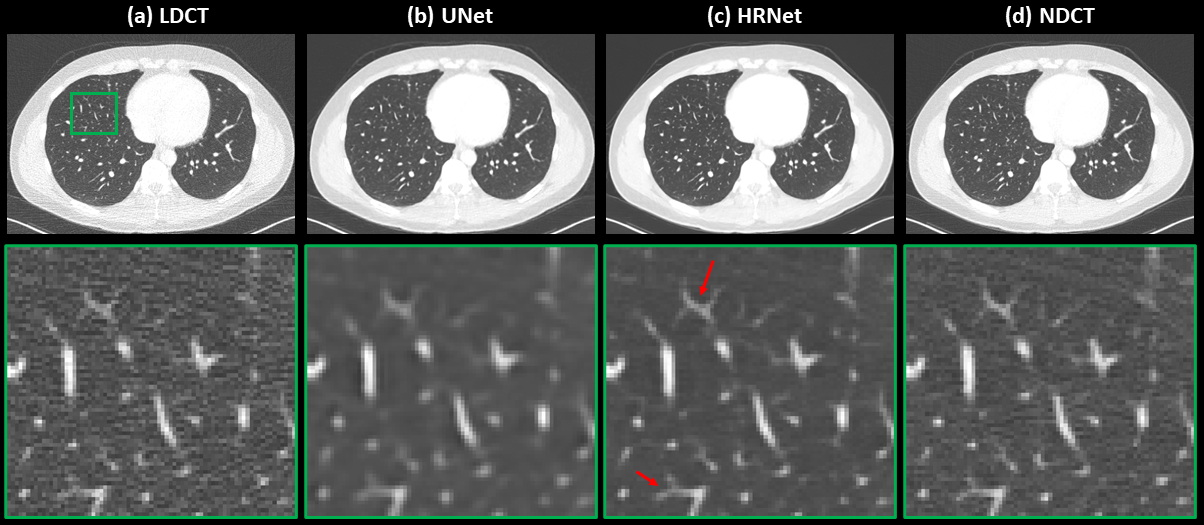}
    \caption{Denoised results for the slice corresponding to the lung region. (a) LDCT, (b) UNet, (c) HRNet, and (d) NDCT. Images in the second row are the zoomed-in views for the contents in the first row, corresponding to the region of interest covered by the solid green box. Display window: [-160, 240] HU.}
    \label{fig:lung}
\end{figure}
Figure~(\ref{fig:lung}) depicts the denoised results for the lung slice. In this case, both the LDCT and the NDCT images can clearly show the locations and the shapes of the lung nodules. After processed by the UNet-based denoiser, the resolution decreases substantially despite the noise is also effectively removed. As for the image generated by the HRNet-based denoiser, we can see that it exhibits much weaker noise while slightly inferior, if not comparable, resolution property than the LDCT image. Indeed, as indicated by the red arrows in Figure~(\ref{fig:lung}), these small lung nodules have clearer structures than those in the LDCT image which are contaminated by the strong quantum noise.

\begin{figure}
    \centering
    \includegraphics[width=\textwidth]{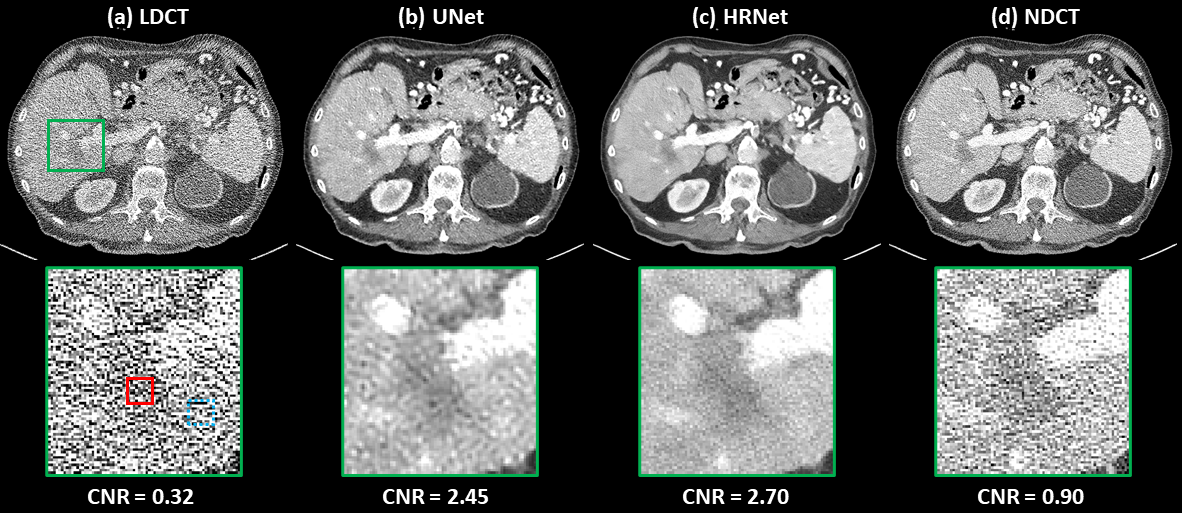}
    \caption{Denoised results for the slice containing a low-contrast lesion, as indicated by the solid green box. (a) LDCT, (b) UNet, (c) HRNet, and (d) NDCT. Images in the second row are the zoomed-in views for the low-contrast lesion. The red solid box and the blue dot box indicate the foreground and the background contents that are used for the CNR calculation, and the associated results are shown below. Display window: [-160, 240] HU.}
    \label{fig:lesion}
\end{figure}
Figure~(\ref{fig:lesion}) compares different denoising results by using the slice that contains a low-contrast lesion, as indicated by the solid green box. Without surprise, it is not easy to delineate this lesion from the surrounding normal tissues by using the noisy LDCT image. This challenge has been greatly alleviated after using the denoisers, as shown by the images in Figures~(\ref{fig:lesion})(b) and~(\ref{fig:lesion})(c). By further inspecting both denoised images, we can observe that the HRNet-based denoiser lead to an image with much more natural noise texture. This can be clearly found from the zoomed-in views in the second row of Figure~(\ref{fig:lesion}). For quantitative comparison, we also calculate the CNRs of this low-contrast lesion on different images, with results of 0.32 (LDCT), 2.45 (UNet), 2.70 (HRNet) and 0.90 (NDCT), further verifying the superior denoising performance of the introduced HRNet-based denoiser.

\begin{table*}[]
    \centering
    \resizebox{\textwidth}{!}{
    \begin{tabular}{ccccccc}
    \hline \hline
    & 	&   Figure~(\ref{fig:abdomen}) & Figure~(\ref{fig:liver}) & Figure~(\ref{fig:lung}) & Figure~(\ref{fig:lesion}) & Dataset \\
    \hline
    RMSE (HU) & LDCT&	115.49&	124.67&	120.90&	164.52&	113.80 \\
    &UNet&	61.08&	63.84&	66.65&	82.32&	59.87 \\
    &HRNet&	56.10&	60.55&	61.01&	76.94&	55.24 \\
    \hline
    SSIM&	LDCT&	0.571&	0.516&	0.588&	0.451&	0.550 \\
    &	UNet&	0.734&	0.666&	0.727&	0.584&	0.712 \\
    &	HRNet&	0.766&	0.702&	0.761&	0.620&	0.745 \\
    \hline \hline
    \end{tabular}}
    \caption{Quantitative results in terms of RMSE and SSIM for the four showcases in Figures~(\ref{fig:abdomen})-(\ref{fig:lesion}) as well as the whole validation dataset. The RMSE has an unit of HU. }
    \label{tab:metrics}
\end{table*}

\begin{figure}
    \centering
    \includegraphics[width=\textwidth]{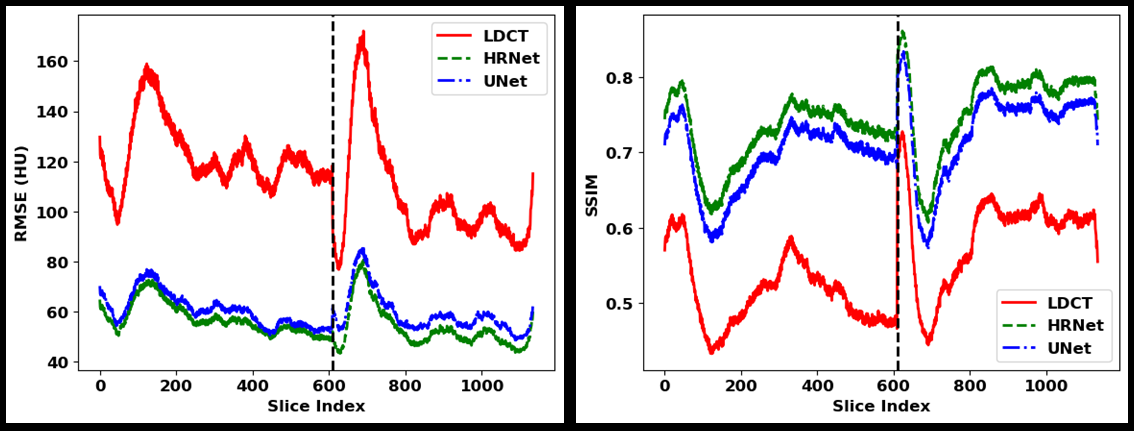}
    \caption{Quantitative results for each slice in the validation dataset. The x-axes are the slice index, the y-axes are the RMSE (left figure) and the SSIM (right figure). The solid red, dash green, dotted dash blue curves are with respect to the LDCT, the HRNet method, and the UNet methods. The slices of the two patients in the validation dataset are separated by the dash black vertical line.}
    \label{fig:metrics}
\end{figure}
To further quantify the denoising performance of both denoisers, we also calculate the RMSE and SSIM values for each showcase in Figures~(\ref{fig:abdomen}) to~(\ref{fig:lesion}) that has the referenced NDCT image, as listed in Table~(\ref{tab:metrics}). It can be seen that both denoisers can dramatically reduce the RMSE values, while the HRNet-based denoiser can lead to an extra improvement than the UNet-based denoiser. Similar findings can also be observed from the SSIM-based metrics. To have a more comprehensive comparison, we calculate and plot the RMSE and SSIM values for all the slices in the validation dataset, as demonstrated in Figure~(\ref{fig:metrics}). Examination of this plot reveals that the introduced HRNet-based denoiser outperforms the UNet-based denoise in almost all the evaluated slices, despite both denoisers can substantially enhance the image quality. As an overall quantitative comparison, we also compute the averaged RMSE and SSIM values for the whole validation dataset. As tabulated in Table~(\ref{tab:metrics}), the RMSE is decreased from 113.80 HU (LDCT) to 59.87 HU if applying the UNet-based denoiser, and is further decreased into 55.24 HU if one uses the HRNet-based denoiser. Again, the SSIM-based metrics also validate the enhancement of image quality, i.e., increased from 0.550 (LDCT) to 0.712 (UNet) and 0.745 (HRNet). 

\begin{figure}
    \centering
    \includegraphics[width=\textwidth]{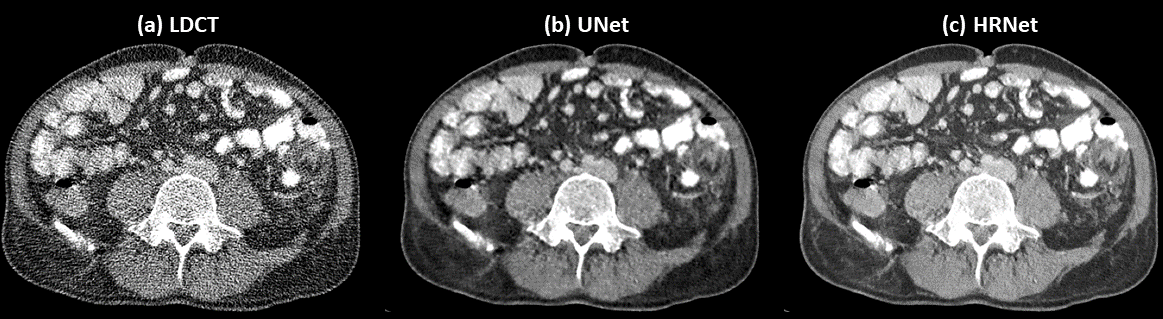}
    \caption{Denoised results for the slice in the testing dataset. (a) LDCT, (b) UNet, (c) HRNet. Display window: [-160, 240] HU.}
    \label{fig:testing}
\end{figure}
The denoising results for a slice in the testing dataset are demonstrated in Figure~(\ref{fig:testing}). Apparently, the HRNet-based denoiser delivers an image with sharper edges while comparable, if not weaker, noise level compared to the UNet-based denoiser. 

\begin{figure}
    \centering
    \includegraphics[width=\textwidth]{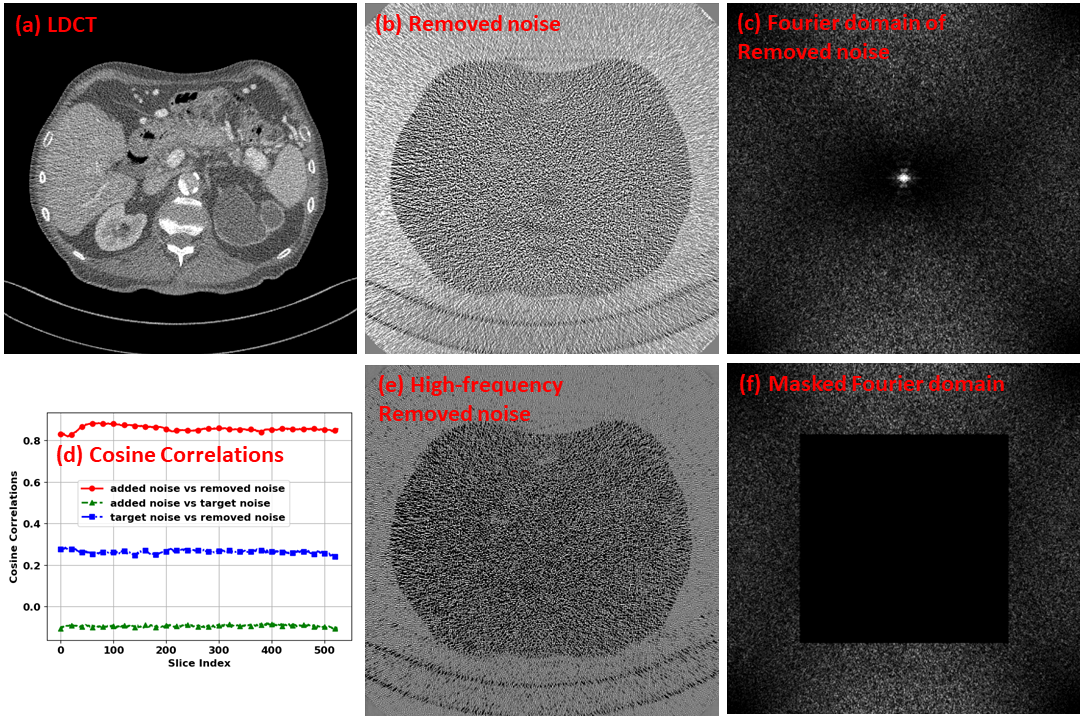}
    \caption{Noise analysis result. (a) LDCT, (b) removed noise by the HRNet-based denoiser, (c) Fourier domain of the removed noise, (d) cosine correlations among the added noise, the removed noise and the target noise, (e) the high-frequency components of the removed noise, (f) masked Fourier domain. The display window for image (a) is [-160, 240] HU, for the images (b) and (e) are [-50, 50] HU, and for the images (c) and (f) are [${10}^4$, ${10}^5$].}
    \label{fig:noise}
\end{figure}
The noise analysis result is shown in Figure~(\ref{fig:noise}). We can observe that the difference image (Figure~(\ref{fig:noise})(b)) between the input and the denoised images almost only contains noise, suggesting that the introduced HRNet-based denoiser can preserve the structural details very well. The corresponding Fourier domain depicted in Figure~(\ref{fig:noise})(c) suggests that the noise is dominant by both the low-frequency and high-frequency components while the middle-frequency noise is effectively removed. After masking out the low-frequency part (Figure~(\ref{fig:noise})(d)) and inversely transforming back to the image domain (Figure~(\ref{fig:noise})(e)), the noise now only contains the high-frequency component, which can be clearly observed by comparing Figures~(\ref{fig:noise})(b) and~(\ref{fig:noise})(e). Besides, since the low-frequency component is excluded, the noise power in Figure~(\ref{fig:noise})(e) is lower than that in Figure~(\ref{fig:noise})(b). The cosine correlations among different noise components are plotted in Figure~(\ref{fig:noise})(d). We can find that the target noise and the extra added noise is almost orthogonal, with a cosine correlation of around -0.08 that is close to zero. The correlation between the removed and the extra added noise is around 0.9, while it is around 0.3 for the correlation between the removed noise and the target noise. These two values indicate that most of the removed noise comes from the extra added noise, while the denoiser also remove some noise from the target image. The projection length from the removed noise to the extra added noise is calculated to be 78.46\%, while the projection length from the removed noise to the target noise is calculated to be 61.31\%. After we subtract both the projected noise from the removed noise, the rest noise is computed to have an energy of 11.37\% of all the removed noise.

\section{Discussions and Conclusions}
The goal of any CT denoisers is to suppress the noise as much as possible and to preserve the anatomical details as well as possible. Deep learning-based denoisers are proved to very effective for noise suppression by automatically extracting the image features, leading to state-of-the-art denoising performance. However, the feature quality highly depends on the model architecture. As for the denoising task, both the low-level and the high-level features are important. The former is of importance for detail-preservation, while the latter is essential for effective noise suppression by using context information from a large scale. The encoder-decoder architecture is efficient for high-level feature extraction, however, it lacks low-level information. The UNet can improve the low-level feature quality to some extent by using skip connections, but still cannot provide a high low-level feature quality that can be used to faithfully restore the fine details. This probably explain why the UNet-based denoiser leads to oversmoothed structures despite it can suppress the noise effectively. By contrast, the introduced HRNet can generate both high-quality low-level and high-level features by using different branches to extract the features from different levels, and also allowing the different level features to be fused together. Consequently, experimental results verified the superiority of this architecture design, showing that HRNet can effectively remove the noise while also preserve the fine anatomical structures very well.

We note that for some cases, the results associated with the HRNet are even better than the NDCT, such as for the low-contrast lesion detection task shown in Figure~(\ref{fig:lesion}). This might be due to that we not only remove the added simulated noise but also remove the noise inherited from the target NDCT image. The noise analysis result demonstrated above can support our hypothesis, with the observation that the HRNet-based denoiser can remove 61.31\% of the target noise from the NDCT image. It should be noted that we are not claiming that the HRNet-based denoiser can deliver an image whose quality surpasses the quality of NDCT. Actually, when the task is to visualize the high-contrast details which are robust to the noise while sensitive to the resolution, such as the lung nodules shown in Figure~(\ref{fig:lung}), the structures associated with the HRNet are slightly oversmoothed compared to the NDCT.

It is well-known that the data-driven deep learning model may suffer from the model generalizability problem when there exists distribution gap between the training and the testing environments. In this work, the models were only tested on the simulated datasets which have similar data distribution as the training dataset if ignoring the potential difference caused by different reconstruction parameters. More realistic testing datasets are required for further performance evaluation before the clinical translation of this model.

In summary, in this work, we introduced a HRNet-based denoiser to enhance the image quality of LDCT images. Benefiting from both the high-quality low-level and high-level features extracted by the HRNet, it can deliver an enhanced image with effectively suppressed noise and well preserved details. Compared to the UNet-based denoiser, the HRNet can produce an image with higher resolution. Quantitative experiments showed that the introduced HRNet-based denoiser can improve the RMSE/SSIM values from 113.80/59.87 (LDCT) to 55.24/0.745, and also outperforms than the UNet-based denoiser whose values are 59.87/0.712.

\section*{Data Availability}
The data used in this paper can be publicly accessed from the official website (https://www.aapm.org/grandchallenge/lowdosect/) of the AAPM Low Dose CT Challenge.

\section*{Acknowledgement}
We would like to thank Varian Medical Systems Inc. for supporting this study and Dr. Jonathan Feinberg for editing the manuscript. We also would like to thank Dr. Ge Wang from Rensselaer Polytechnic Institute for his constructive discussions and comments.

\bibliographystyle{IEEEtran}
\bibliography{./HRNet}
\end{document}